\documentclass{elsart} %
\usepackage{amssymb}
\usepackage{amsmath}

\newcommand{\be}{\begin{equation}}
\newcommand{\ee}{\end{equation}}

\newcommand{\tr}{{\mathrm{Tr}}}

\begin{document}
\begin{frontmatter}

\title{
Information measures based on Tsallis' entropy and geometric considerations
for thermodynamic systems }

\author[laplata]{M.\ Portesi},
\author[laplata]{A.\ Plastino}, and
\author[laplata,antofagasta]{F.\ Pennini}

\address[laplata]{Instituto de F\'{\i}sica La Plata
(CONICET--UNLP), \\
Dpto.\ de F\'{\i}sica, Fac.\ de Ciencias Exactas,
Universidad Nacional de La Plata \\
C.C.~67, 1900 La Plata, Argentina}

\address[antofagasta]{Departamento de F\'{\i}sica, Universidad Cat\'olica del
Norte,\\
Casilla 1280, Antofagasta, Chile}



\begin{abstract}

An analysis of the thermodynamic behavior of quantum systems can be
performed from a geometrical perspective investigating the structure of the
state space. We have developed such an analysis for nonextensive
thermostatistical frameworks, making use of the $q$-divergence derived from
Tsallis' entropy. Generalized expressions for operator variance and
covariance are considered, in terms of which the fundamental tensor is
given.



\end{abstract}

\maketitle
\end{frontmatter}

\section{Introduction}
\label{section_introduction}

The geometrization of thermodynamics and statistical mechanics has
been the subject of many studies during the last few decades.
Among the collection of works related with this topic, different
approaches have been followed. For instance, Weinhold
\cite{w_jcp1975} considered the thermodynamic surface given by the
fundamental relation $U=U(\{X_i\})$ in the $(r+1)$-dimensional
Gibbs space (with coordinates labelled by $U,X_1,\ldots,X_r$), and
obtained the components of the metric tensor of that space as the
second derivatives of the internal energy $U$ with respect to each
pair of the $r$ extensive parameters $X_i$. Ruppeiner
\cite{r_pra20rmp67} focuses attention on fluctuations of the
thermodynamic magnitudes and obtains the metric tensor via second
moments of the fluctuations. The two ensuing metrics have been
proven by Mruga{\l}a to be equivalent \cite{m_p125a}. Another
statistical path for reaching a Riemannian metric
\cite{lc_sa191poole} in the space of thermodynamic parameters is
that originated in the works by Rao \cite{r_bcms37} and Amari
(see, for instance, \cite{amarian_2000}) in the field of statistical
mathematics, and also by Ingarden \cite{i_tns}, Janyszek
\cite{j_rmp24}, and other authors in the field of thermodynamics
and statistical mechanics. The concomitant information-theoretic
approach is based on the concept of relative entropy and given in
terms of the Boltzmann--Gibbs--Shannon entropy. The ensuing
formalism has been applied to a number of model systems, from the
ideal and van der Waals gases to the Ising and other magnetic
models. In these applications it has been seen that the scalar
curvature $R$ of the space can represent a measure of the
thermodynamic stability of the system, and a useful quantity to
characterize phase transitions (typically, for non-interacting
models one obtains $R=0$, i.e.\ a flat geometry, while $R$
diverges at the critical point for interacting systems). There are
some recent efforts in the field of information geometry related
with the generalized, nonextensive formulation of statistical
mechanics \cite{t_jsp52}. Among these studies one finds the
analysis by Abe \cite{a_pre68} of the geometry of escort
distributions, and the contributions by Amari, Nagaoka and
coworkers (\cite{amarian_2000}, among others) in connection with
the geometrical structure in the manifold of probability
distributions. A new geometrical approach to thermo-statistical
mechanics is introduced in~\cite{tb_0203536}, where the relevance
of the approach within the contexts of nonextensive statistical
thermodynamics is analyzed, showing that Riemannian geometry
concepts yield a powerful tool. More recently, Naudts
\cite{n_osid12} studies escort density operators and generalized
Fisher information measures. %

Here, our aim is to discuss in some detail the generalization of
the geometrical approach to statistical physics. For that purpose
we appeal to a generalized form for the entropy, as given by
Tsallis \cite{t_jsp52}. The $q$-entropy is employed in order to
define an information measure from which we can derive the
metrical structure of the parameters' space. The generalization of
the definitions of variance and covariance for quantum operators
in the context of the so-called OLM version \cite{mnpp_pa286} of
nonextensive statistical mechanics is developed. The metric is
finally expressed in terms of generalized fluctuations.

\section{OLM density operator}
\label{section_olm}

For a given quantum mechanical system, the density operator that maximizes
Tsallis' nonextensive $q$-entropy \cite{t_jsp52} \ $S_q\equiv k_B (1-{\rm
Tr}\,\hat\rho^q)/(q-1)$ \ (with $q\in\mathbb{R}^+$ and $k_B\equiv 1$) is
written, within the optimized Lagrange multipliers´ (OLM)
formalism~\cite{mnpp_pa286}, as \ $\hat\rho={\bar Z_q}^{\ -1} \;
e_q\!\left(-\sum_{i=1}^r \lambda_i (\hat F_i-m_i)\right)$. Here, the
generalized expectation values of $r$ quantum operators $\{\hat
F_1,\ldots,\hat F_r\}$ are considered to be known as prior information; they
are given by \ $m_i=\langle\hat F_i\rangle_q=\tr\left(\hat\rho^q \hat
F_i\right)/\tr \hat{\rho}^q$, $i=1,\ldots,r$. The parameters
$\{\lambda_1,\ldots,\lambda_r\}$ refer to the set of Lagrange multipliers
that fit those restrictions in the procedure of constrained extremization of
$S_q$ when one is working {\it within the OLM formalism}, i.e.\ the
restrictions are rewritten as \ ${\rm Tr}\,\hat\rho^q (\hat F_i-m_i)=0$. It
has been established~\cite{ampp_pla281a_pa300} that $\{\lambda_i\}$
correspond to the {\it physical}\ intensive parameters. In this OLM
framework, the pseudo-partition function \ $\bar Z_q\equiv{\rm
Tr}\,e_q\!\left(-\sum \lambda_i (\hat F_i-m_i)\right)$ \ is such that the
density operator is normalized, i.e.\ ${\rm Tr}\,\hat\rho=1$. Notice that
the Lagrange multiplier associated with the normalization condition is not
written explicitly in the equilibrium density matrix, instead we chose to
introduce the $q$-partition function $\bar Z_q$ in the expression for
$\hat\rho$. In all these expressions, $e_q(x)$ stands for the
$q$-exponential function: \ $e_q(x)\equiv[1+(1-q)x]_+^{1/(1-q)}$, with
$[X]_+=\max\{0,X\}$. The extensive limit corresponds to the situation
$q\rightarrow 1$, and then $q-1$ is a measure of the degree on entropy
nonextensivity.

\section{Generalized variances}
\label{section_qvariances}

We begin by providing a natural $q$-generalization of the
variance of an operator and of the covariance between two
operators. For this purpose we address the relation between
fluctuation and response in nonextensive settings
\cite{sm_pc2004}. We need then to compute the derivative of each
mean value with respect to every Lagrange multiplier, which poses
a rather intricate problem. A linear system of $r$ coupled
equations is to be faced for each $\lambda$. This system can be
solved and, after some manipulations, the following result is
reached
{\small
\begin{eqnarray}
\frac{\partial m_i}{\partial\lambda_j}
= -q \, \bar Z_q^{q-1} \left(1-q \bar
Z_q^{q-1}\sum_{l=1}^r\lambda_l \langle\hat\rho^{q-1}\delta_q\hat
F_l\rangle_q\right)^{-1} \bigg[
\langle\hat\rho^{q-1}\,\delta_q\hat F_i\,\delta_q\hat F_j\rangle_q
- q \bar Z_q^{q-1}\times &&
\nonumber\\
\times\bigg( \langle\hat\rho^{q-1}\,\delta_q\hat F_i\,\delta_q\hat
F_j\rangle_q \,
\sum_{l=1}^r\lambda_l\langle\hat\rho^{q-1}\delta_q\hat
F_l\rangle_q -\langle\hat\rho^{q-1}\delta_q\hat F_i\rangle_q
\sum_{l=1}^r\lambda_l \langle\hat\rho^{q-1}\,\delta_q\hat
F_j\,\delta_q\hat F_l\rangle_q \bigg)\bigg] && \nonumber
\end{eqnarray}
} where $\delta_q\hat F_i\equiv\hat F_i-m_i$ are the generalized deviation
operators, and $\hat\rho$ is the OLM density matrix. For the sake of
simplicity, we have considered here a set of $r$ {\it commuting} operators.
Let us stress that the derivative with respect to a given $\lambda_j$ is
done keeping all other $\lambda_{j'} \, (j'\neq j)$ fixed. After writing
down the above equation it seems to us advantageous to advance the following
definitions for $q$-generalized deviations, covariances, and squared
variances or dispersions, respectively
\begin{eqnarray}
(\delta_q F_i) & \equiv & \langle\hat\rho^{q-1}\,\delta_q\hat
F_i\rangle_q
\label{qdev}\\
C_q(\hat F_i,\hat F_j)  =  \langle\hat\rho^{q-1}\delta_q\hat F_i
\,\delta_q\hat F_j\rangle_q \qquad & \textrm{and} & \qquad (\Delta_q
F_i)^2 \equiv
C_q(\hat F_i,\hat F_i)
\label{qvar}
\end{eqnarray}
These can be interpreted as modified first and second moments of the
corresponding operators. For the sake of completeness, we also define the
generalized correlation coefficient to be \ ${\mathcal C}_q(\hat F_i,\hat
F_j) \equiv {C_q(\hat F_i,\hat F_j)}/\left({\Delta_q F_i\;\Delta_q
F_j}\right)$, that equals 1 whenever $i=j$ for {\it arbitrary} values of
$q$. Regarding the first definition, Eq.~(\ref{qdev}), the generalized
expectation value in the r.h.s.\ will {\it not} be equal to zero in general
--notice that, as it is given here, it is not the $q$-mean value of the
$q$-deviation operator--; however $\lim_{q\rightarrow 1}(\delta_q F_i)=0$.
The expressions in Eq.~(\ref{qvar}) for the second moments differ from those
given in~\cite{a_pa269} in a factor $\hat\rho^{q-1}$ inside the
$q$-expectation values, that can be recast as \ $\bar
Z_q^{1-q}\,[1-(1-q)\sum\lambda_l\,\delta_q\hat F_l]_+^{-1}$ \ (a similar
factor has also been found in the computation of the generalized specific
heat for an ideal Fermi gas \cite{mppp_pa332}). Typical of the nonextensive
statistical formalism is the emergence of {\it correlations} among different
observables --induced by the nature of the $q$-statistics-- with one
quantity depending on {\it all} other ones (this has also been discussed for
the occupation numbers in fermionic systems \cite{mppp_pa332}).
Nevertheless, one always finds the correct uncorrelated limit for
$q\rightarrow 1$. Due to the presence of the density-dependent factor in
Eq.~(\ref{qvar}), the evaluation of the $q$-variances and $q$-covariances
involves not only quadratic terms with $\hat F_i \hat F_j$ (as
required in the extensive limit) but the computation of $(r+1)(r+2)/2$
traces, apart from ${\rm Tr}\,\hat\rho^q$, for a complete description of the
correlations for a given problem. Indeed we can write
\begin{eqnarray}
(\Delta_q F_i)^2 & = &
\langle\hat\rho^{q-1}\hat F_i^2\rangle_q - 2 m_i \,
\langle\hat\rho^{q-1}\hat F_i\rangle_q + m_i^2 \,
\langle\hat\rho^{q-1}\rangle_q
\label{qvar3}
\\
C_q(\hat F_i,\hat F_j) & = &
\langle\hat\rho^{q-1}\hat F_i \hat F_j\rangle_q
- m_i \, \langle\hat\rho^{q-1}\hat F_j\rangle_q
- m_j \, \langle\hat\rho^{q-1}\hat F_i\rangle_q
+ m_i m_j \, \langle\hat\rho^{q-1}\rangle_q
\label{qcovar4}
\end{eqnarray}
A bit of additional algebra finally yields one of our important
results, namely,
\begin{equation}
\left.\frac{\partial
m_i}{\partial\lambda_j}\right|_{\{\lambda_{j'\neq j}\}}
 = -q\bar Z_q^{q-1}\left(C_q(\hat F_i,\hat F_j)+
\frac{q\bar Z_q^{q-1}\,(\delta_q F_i)\,\sum\lambda_l\,C_q(\hat F_j,\hat F_l)}
{1-q\bar Z_q^{q-1}\,\sum\lambda_l\,(\delta_q F_l)}
\right)
\end{equation}

\section{The fundamental tensor}
\label{section_fundamentaltensor}

The fundamental tensor of the space of parameters, a key ingredient in the
geometric approach to thermostatistics, can be interpreted in terms of
thermodynamic fluctuations. Indeed, in the case of classical systems or
commuting operators, the metric tensor derived within a standard treatment
is equal to the covariance or second moment: \ $g_{ij}^{(1)}=\langle\,(\hat
F_i-\langle\hat F_i\rangle) (\hat F_j-\langle\hat
F_j\rangle)\,\rangle=C_1(\hat F_i,\hat F_j)$. \ For non-commuting operators,
an integral expression for the covariances has been
introduced~\cite{jm_pra39}, based on the connection with the metric tensor.
We present in this section the formalism leading to analogous results within
generalized statistical contexts.

A quantum state described by the density operator
$\hat\rho(\lambda_1,\ldots,\lambda_r)$ can be represented in the
$r$-dimensional space of parameters. The {\it information distance} between
two normalized states can be given in terms of the symmetrized form of the
relative entropy, which in a nonextensive context has been defined as
\cite{a_pa344} \
$K_q(\hat\rho\|\hat\sigma)=\tr\hat\rho^q(\ln_q\hat\rho-\ln_q\hat\sigma)$,
where $\ln_q(x)$ stands for the inverse function of $e_q(x)$. We compute
then the symmetric information measure for two neighbor density matrices,
$\hat\rho(\{\lambda\})$ and $\hat\rho(\{\lambda+\partial\lambda\})$, and
make an expansion around $\{\lambda\}$. The first non-vanishing contribution
is the second order one, $\sum\partial\lambda_i\partial\lambda_j
\,q\tr(\hat\rho^q)\langle\hat\rho^{-q-1}\partial_j\hat\rho\,
\partial_i\hat\rho\rangle_q$ \, (where $\partial_iX\equiv\partial X/\partial\lambda_i$),
which finally gives us the $q$-metric tensor as
\begin{eqnarray}
g^{(q)}_{ij} = q \, {\bar Z_q}^{\,q-1} \, && \bigg[
C_q(\hat F_i,\hat F_j) -
\nonumber\\
&& - \partial_i\ln\bar Z_q \, (\delta_q F_j) -
\partial_j\ln\bar Z_q \, (\delta_q F_i) +
\nonumber\\
&&
+\partial_i\ln\bar Z_q \, \partial_j\ln\bar Z_q \left(
\langle\hat\rho^{q-1}\rangle_q-\bar Z_q^{\,1-q}\right) \bigg]
\label{gqijfluc}
\end{eqnarray}
Given in this shape, the evaluation of the generalized metric tensor for a
given system requires knowledge of the pseudo-partition function and its
logarithmic derivatives with respect to the Lagrange parameters, and also
the generalized fluctuations.

\hfill


Summing up, we have discussed appropriate definitions of $q$-variance and
$q$-covariance for quantum operators in a generalized thermostatistical
framework characterized by the nonextensivity index $q$, along the paths of
Ref.~\cite{sm_pc2004}. Previous related literature is based on the
generalized definitions given in~\cite{a_pa269}. Then, we have found the
fundamental tensor of the space of thermodynamic parameters within a
nonextensive statistical framework, in terms of quantum fluctuations.
Application of these ideas to certain physical systems may contribute to
characterize its thermodynamic behavior. In this sense we expect that the
geometric analysis of the model, when performed in a generalized context
with a value of $q$ different from 1, may exhibit in a more clear way the
critical regions.

\section{Acknowledgments}

Financial support from CONICET and UNLP, Argentina, is
acknowledged. MP also acknowledges ANPCyT (PICT
No.~03-11903/2002), Argentina.



\end{document}